# Transmitting FRET signals to nerve cells

Jakub Kmiecik
Department of Telecommunications,
AGH University of Science and Technology,
30-059 Krakow, Poland,
jkmiecik@agh.edu.pl

Pawel Kulakowski
Department of Telecommunications,
AGH University of Science and Technology,
30-059 Krakow, Poland,
kulakowski@kt.agh.edu.pl

Krzysztof Wojcik
Department of Medicine,
Jagiellonian University Medical College,
31-008 Krakow, Poland,
krzysztof.wojcik@uj.edu.pl

Andrzej Jajszczyk
Department of Telecommunications,
AGH University of Science and Technology,
30-059 Krakow, Poland,
jajszczyk@kt.agh.edu.pl

## Abstract

This paper is concerned with a novel method allowing communication between FRET nanonetworks and nerve cells. It is focused on two system components: fluorophores and channelrhodopsins which serve as transmitters and receivers, respectively. Channelrhodopsins are used here also as a FRET signal–to–voltage converter. The trade-off between throughput and bit error rate is also investigated.

## CSS Concepts

• **Applied computing~Biological networks**
• *Networks~Network performance analysis* • *Computing methodologies~Molecular simulation*

## Introduction

Single nanomachines according to their size will be able to perform only simple tasks. Thus, a need emerges to connect them so they can cooperate to perform complex operations. Many nanocommunication techniques have been already proposed but their common drawbacks are large propagation delays and slow transmission rates. Förster Resonance Energy Transfer (FRET) is a physical phenomenon, in which one excited fluorescent molecule (donor) irradiatively passes its energy to another fluorescent molecule (acceptor). The average FRET delay is few ns and the whole energy transfer process lasts no more than 40 ns, what results in even 25 Mbps throughput [1]. The probability of successful energy passing is strongly dependent on the distance *r* between donor and acceptor and can be expressed as [2]:

$$E = \frac{R_0^6}{r^6 + R_0^6}$$

In above formula, $R_0$ is a constant characteristic for a particular donor – acceptor pair and is called Förster distance. In order to achieve high values of Förster distance, the donor emission spectrum and acceptor absorption spectrum have to overlap each other. Other factors that affect this parameter are: the donor quantum yield, relative orientation between donor and acceptor and medium. It typically ranges from 10 to 100 Å.

For communication purposes the donor serves as a transmitter and the acceptor as a receiver. Single bits are sent over communication channel utilizing ON-OFF modulation; sending bit '1' is done by donor excitation and energy transfer to acceptor and when sending bit '0' no donor is excited and no energy transfer occurs. Thus, '0' bit transmission is always correct and bit error rate (BER) can be calculated as:

$$BER = 0.5(1 - E)$$

As FRET communication has been investigated, some limitations have been revealed. One of them is its high BER, which has been dealt by utilizing MIMO-FRET [3]. In this concept, on the transmitter side there are multiple donors and analogously, on the receiver side there are multiple acceptors. This technique allowed to achieve BER about 10$^{-3}$.

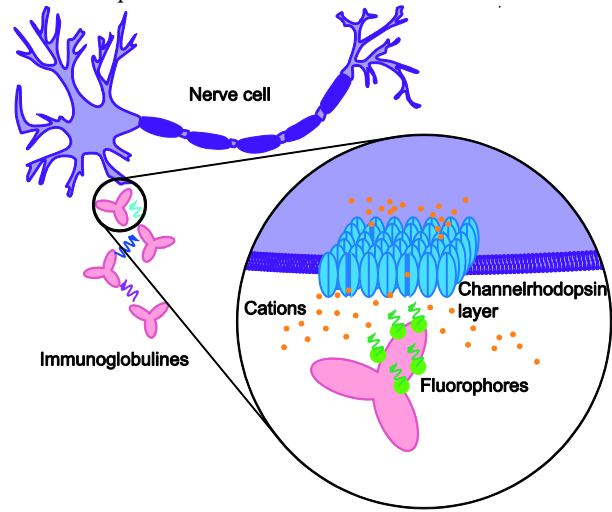

Fig. 1. The nanocommunication system linking a FRET network and a nerve cell.

## Connection between FRET network and neurons

Presented FRET interface converts signals from FRET network to electrical signals. Following system components can be distinguished:

• Channelrhodopsins (ChRs)–proteins that perform a function of ion channels, i.e. are able to pass cations through a channel pore. The heart of a channelrhodopsin is a retinal complex, which serves as a FRET signal receiver. When a retinal becomes excited, the pore opens for at least 10 ms and during this period about $3 \times 10^2$ ions can pass through, then the ChR remains inactive for about 50 ms [4]. This long inactivity time comparing to the FRET process duration would cause a bottleneck in the system, thus multiple ChRs have been used. They have been arranged in a square matrix which can be embedded into nerve cell membrane (see Fig. 1) replacing neuroreceptors in nerve cell [5].

- Fluorophores – molecules that transmit FRET signals to retinal complexes and also act as sinks for the rest of the FRET network. They reside on immunoglobulin G molecules.
- Immunoglobulin G (IgG) – an antibody molecule that serves as a carrier for up to 9 fluorophores [6]. IgG is immobilized via a Fc ligand–Fc receptor bond which is linked by a tunable length amino acids sequence to the nerve cell membrane [7].

The principle of operation of the system remains the same as in MIMO-FRET networks. When sending a bit '1', excited fluorophores located on the IgG pass their energy to retinal complexes in ChRs. Successful bit transmission is considered when at least one retinal complex gets excited and a pore channel opens. Opened channels cause cations to flow, what results in creating a potential difference between the nerve cell and the environment what leads to propagation of an action potential in a neuron. By sending a bit '0' no energy is transferred, so there is no voltage change.

## Simulation scenario and results

Simulations were conducted in MATLAB using suitable molecular models of system components. There was 1IGT model used for IgG, 3UG9 model for ChRs (both from Protein Data Bank) and Atto 390 NHS ester model (from Pubchem database). The choice of fluorophore was determined by its high $R_0$ = 52 Å. In the IgG, all possible 11 locations where a fluorophore can be attached were determined by finding solvent accessible lysine side chains [8]. For each simulation run fluorophores positions and orientations were randomly chosen taking into account peptide bond geometry and steric effects. Later, the optimal spatial position of IgG relative to the ChR matrix was determined, i.e. the position where the FRET efficiency between all 11 possible fluorophore positions and all retinal complexes was the highest. The IgG was distant from ChR layer by 20 Å – this distance was selected so the IgG was close to the ChR layer but ions could still freely diffuse through the channel pores. The ChR matrix size was chosen as 7x7, as the larger matrices had not improved the transmission efficiency (the border retinals had been too far from the fluorophores). The simulations were run 1000 times per each considered transmission rate. In each simulation run 1000 bits were sent.

During the simulations, the BER of such a MIMO-FRET communication was investigated for different transmission rates (see Fig. 2). If the throughput was low, below 17 b/s (what is one bit per ChR activity cycle), the BER was constant and was equal to $10^{-3}$ for 9 fluorophores. However, with the higher transmission rate, the BER started to increase, as the ChRs were still in their inactive phase when another bit '1' was transmitted. These results show a kind of trade-off between the transmission rate and quality

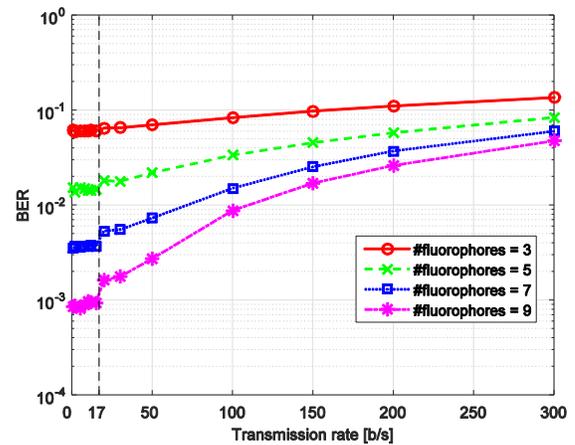

Fig. 2. Bit error rate as a function of transmission rate for communication between fluorophores and channelrhodopsins.

## Conclusions

In this paper, a novel communication scheme for nanonetworks was introduced. It was presented how FRET-based nanonetworks can send their signals to neurons, i.e. nerve cells. The possible transmission throughputs and achievable bit error rates were shown. This new interface between FRET-based and neural networks creates also a huge potential for numerous medical applications where nanonetworks could stimulate the human neural system.

## Acknowledgement

This work has been performed under the contracts 11.11.230.018 and 15.11.230.266.